\documentclass[aps,reprint,pre,10pt,superscriptaddress,nobibnotes,showpacs]{revtex4-1}
\usepackage{amsfonts}
\usepackage{amsmath}
\usepackage{amssymb}
\usepackage{graphicx}

\begin{document}

\title{Upper-hybrid wave driven Alfv\'{e}nic turbulence in magnetized dusty
plasmas}
\author{A. P. Misra}
\email{apmisra@visva-bharati.ac.in}
\affiliation{Department of Physics, Ume{\aa } University, SE-901 87 Ume{\aa}, Sweden}
\author{S. Banerjee}
\affiliation{Department of Mathematics, Politecnico di Torino, Turin, Italy}
\begin{abstract}
The nonlinear dynamics of coupled electrostatic upper-hybrid (UH)
and Alfv\'{e}n waves (AWs) is revisited in a magnetized electron-ion plasma
with charged dust impurities. A pair of nonlinear equations  that describe the interaction of UH wave
envelopes (including the relativistic electron mass increase) and the
density as well as the compressional magnetic field perturbations
associated with the AWs is solved numerically to show that many coherent
solitary patterns can be excited and saturated due to modulational
instability  of unstable UH waves. The evolution
of these solitary patterns is also shown to appear in the states of spatiotemporal
coherence, temporal as well as spatiotemporal chaos due to collision
and fusion among the patterns in stochastic motion. Furthermore, these spatiotemporal features
are demonstrated by the analysis of wavelet power spectra. It is found
that a redistribution of wave energy takes place to higher harmonic
modes with small wavelengths which, in turn, results into the onset
of Alfv\'{e}nic turbulence in dusty magnetoplasmas. Such a scenario
can occur in the vicinity of Saturn's magnetosphere as many electrostatic
solitary  structures have been observed there by the Cassini spacecraft.
 
\end{abstract}
\received{31 October 2010}
\revised{08 December 2010}
\maketitle
\startpage{1} \endpage{102}

The nonlinear interaction of high-frequency (hf) electrostatic upper-hybrid
(UH) waves (ES-UHWs) and the low-frequency (lf) electrostatic or electromagnetic
waves in a magnetized dusty plasma is a topic of current research
(see, e.g., Refs. \cite{ES-UH-Envelope,JPP2007,Upper-hybrid-Simulation,UH-Review,UH-Kaufman,Modified-Alfven-Wave})
as those give rise to a great variety of nonlinear effects including
parametric instabilities \cite{Parametric-Instability}, decay \cite{Decay-Stenflo,Decay-Goodman}
as well as modulational interactions \cite{ES-UH-Envelope,JPP2007,Upper-hybrid-Simulation,UH-Review,UH-Kaufman,Modified-Alfven-Wave}.
The latter can be responsible for the evolution of UH envelope solitons
\cite{ES-UH-Envelope,JPP2007,Upper-hybrid-Simulation} and UH wave
collapses \cite{Wave-Collapse,Wave-Collapse-Stenflo} which have been
observed in laboratory and space plasmas \cite{Wave-Collapse,experiments}.
Moreover, the formation of such envelope solitons through the nonlinear
interaction of coupled hf and lf waves is one of the most interesting
features in modern plasma physics in the context of plasma turbulence
\cite{STC-He,STC-QZEs,STC-Rizzato,STC-ZEs,TC-QZEs}, plasma heating and
particle acceleration \cite{Electron-Acceleration-Turbulence}. Furthermore,
the pattern formation and the existence of spatiotemporal chaos (STC)
characterized by its extensive and irregular pattern dynamics in both
space and time in nonlinear dynamical systems have received renewed
interests (see, e.g. Refs. \cite{STC-He,STC-QZEs,STC-ZEs,TC-PRL}).

When the electric field intensity becomes strong and approaches the
modified decay instability threshold, the interaction between hf and
lf waves results to `weak turbulence' in which hf waves are scattered
off lf waves. However, if the electric field is so strong that it
exceeds the modulational instability (MI) threshold, the interaction
is then said to be in `strong turbulence' regime in which transfer
or redistribution of wave energy to higher harmonic modes with small
wavelengths can take place \cite{STC-He,STC-QZEs,STC-ZEs}. In this
context, it is, however, believed that some sort of chaotic process
may be responsible for the transfer of energy from large to small
spatial scales \cite{STC-Rizzato}.

Recently, Shukla \textit{et al.} \cite{JPP2007} have developed the nonlinear
theory of coupled mode interactions between ES-UHWs and modified Alfv\'{e}n
waves (MAWs), and predicted its application to the dusty magnetosphere
of Saturn. However, they extended this work by accounting for the effects of the
relativistic electron mass increase in the UH fields and the density
as well as compressional magnetic field perturbations driven by the
UH ponderomotive force. They  made observations on a new class
of oscillatory MI and spiky UH wave envelopes trapped in the electron
density cavity.

On the other hand, Williams \textit{et al} \cite{Solitary-Wave-Saturn} have
reported observations of electric field solitary structures by Cassini
spacecraft in the vicinity of Saturn's magnetosphere with ambient
magnetic fields $\sim0.1$ to $8000$ nT. They  also reported
a range of the peak-to-peak wave amplitude by the external magnetic
field as well as time durations of the solitary pulses. Moreover,
it has been suggested that the turbulence in the Earth's magnetosheath
may account for the observations of many solitary pulses there \cite{Turbulence-Pickett}.
Also, majority of solitary structures have been found at Europa in
the moon's wake, a region of known turbulence \cite{Turbulence-Kurth}.
It is, therefore,  of natural interest to investigate the
possibility for the occurrence of turbulence in the vicinity of dusty
magnetosphere of Saturn where many ES-UHWs have been observed \cite{Solitary-Wave-Saturn}. We
report here, possibly for the first time, the onset of lf Alfve\'{e}nic turbulence in the
nonlinear interaction of ES-UHWs and MAWs in parameter regimes that are relevant to dusty magnetosphere
of Saturn. We find that redistribution of wave energy indeed takes
place to higher harmonic modes, and the process becomes faster the
smaller is the wave number of modulation at which many unstable modes
are to be excited and saturated by the MI of ES-UHWs.

In what follows, we consider the following nondimensional set of equations
describing the dynamics of coupled ES-UHWs and MAWs \cite{JPP2007}.
\begin{eqnarray}
& & i\left(\partial_{t}+a_{1}\partial_{x}\right)E+a_{2}\partial_{x}^{2}E+a_{3}\left(|E|^{2}-N\right)E=0,\label{eq:b1}\\
& & \left(\partial_{t}^{2}+b_{1}-b_{2}\partial_{x}^{2}\right)N-b_{3}\partial_{x}^{2}|E|^{2}=0,\label{eq:b2}\end{eqnarray}
 where the coefficients are $a_{1}=3k_{0}\alpha V_{Te}/\omega_{H},\: a_{2}=3\alpha\omega_{pi}/2\omega_{H},\: a_{3}=\Omega_{0}^{2}/2\omega_{H}\omega_{pi},\: b_{1}=\Omega_{R}^{2}/\omega_{pi}^{2},\: b_{2}=V_{AR}^{2}/V_{Te}^{2}$ and $b_{3}=m\mu c^{2}/3V_{Te}^{2}$
in which $k_{0}$ is the UH wave number, $\alpha=1/\left(1-3\omega_{c}^{2}/\omega_{p}^{2}\right)>0$
with $\omega_{c(p)}$ denoting the electron cyclotron (plasma) frequency,
$V_{Te}$ is the electron thermal speed, $\omega_{H}=\sqrt{\omega_{p}^{2}+\omega_{c}^{2}}$
 is the UH resonance frequency and $\omega_{pi}$ is the ion plasma
frequency. Also, $\Omega_{0}=\sqrt{\omega_{p}^{2}+2\omega_{c}^{2}}$,
$\Omega_{R}=Z_{d}n_{d0}\omega_{ci}/n_{e0}$ is the Rao cut-off frequency
\cite{Rao-cut-off}, $Z_{d}$ is the number of electrons residing
on a stationary charged dust grain with unperturbed number density
$n_{d0}$, $V_{AR}=\mu B_{0}/\sqrt{4\pi n_{i0}m_{i}}$ is the modified
Alfv\'{e}n speed \cite{Rao-cut-off}, $\mu=n_{i0}/n_{e0}(=1+\delta\equiv1+Z_{d}n_{d0}/n_{e0})$
is the unperturbed ion to electron number density ratio, $m=m_{e}/m_{i}$
is the electron rest to ion mass ratio and $B_{0}$ is the external magnetic
field. Moreover, $E=E_{x}/E_{c}$ is the \emph{$x$}-component of
the nondimensional UH wave electric field with $E_{c}=2m_{e}c\omega_{p}^{2}/\sqrt{3}e\omega_{H}$
and $N=n_{e1}/n_{e0}(=B_{1z}/B_{0}\ll1)$ denoting the normalized
electron number density perturbation (compressional magnetic field)
associated with the MAWs in dusty plasmas \cite{ES-UH-Envelope}. The space and time variables
are normalized respectively by $V_{Te}/\omega_{pi}$ and the ion plasma
period $\omega_{pi}^{-1}$. The fourth and fifth term $\propto a_{3}$
in Eq. (\ref{eq:b1}) arise respectively due to the relativistic electron
mass increase and the density as well as the compressional magnetic
field fluctuations that are driven by the ES-UHW ponderomotive force [For details about Eqs.  (\ref{eq:b1}) and (\ref{eq:b2}) see, e.g. Ref. \cite{JPP2007}].

Now, the dispersion relation for the MI of a constant amplitude UH pump
of the form $E_{0}\exp\left(ik_{0}x-i\omega_{H}t\right)$+complex
conjugate, can readily be obtained. Here we decompose the electric
field $E$ as the sum of the pump and its two sidebands, and obtain
the following dispersion law \cite{JPP2007} (Note that some terms
in Ref. \cite{JPP2007} might be missing or have been simplified somehow).
$\left(\Omega^{2}-\Omega_{AR}^{2}\right)\left[\left(\Omega-KV_{g}\right)^{2}-\Delta\right]=2a_{3}b_{3}K^{2}E_{0}^{2}\Lambda,$
 where $\Delta=\Lambda\left(\Lambda-2a_{3}E_{0}^{2}\right)$, $\Lambda=a_{2}K^{2}-a_{3}E_{0}^{2}$,
$\Omega_{AR}^{2}=b_{1}+b_{2}K^{2},$ $V_{g}=a_{1}$ are in nondimensional
forms and $\Omega$, $K$ are the normalized wave frequency and wave
number of modulation respectively. For the parameter values as relevant
to the Saturn's magnetosphere, namely $n_{e0}=80$ {cm}$^{-3}$,
$\delta=10$, $T_{e}=3\times10^{7}$ K, $B_{0}=1500$ nT \cite{Solitary-Wave-Saturn} 
we find that the coefficients of $\Omega^{3}$ and $\Omega$ become
smaller compared to the other terms. Hence the above dispersion relation  can be simplied to obtain
$\Omega^{2}\approx\frac{1}{2}\left[\Omega_{AR}^{2}-K^{2}V_{g}^{2}+\Delta\right]\pm\frac{1}{2}\left(\left[\Omega_{AR}^{2}+K^{2}V_{g}^{2}-\Delta\right]^{2}+8a_{3}b_{3}K^{2}E_{0}^{2}
\Lambda\right)^{1/2}.$  
 This shows that the ES-UHWs are modulationally unstable when the expression
inside the square root becomes negative (except the case in which
$\Lambda>0$ for $K>\sqrt{a_{3}/a_{2}}E_{0}$ giving a purely damping
mode for the lower sign) for $K<K_{c},$ a critical wave number. We
numerically investigate some domains of $K$ with the parameters as
above for different values of the initial pump for which the MI sets
in. These are namely, $0.23\lesssim K\lesssim0.51$ for $E_{0}=0.2$; $0.47\lesssim K\lesssim0.83$
for $E_{0}=0.3$; $0.74\lesssim K\lesssim1.1$ for $E_{0}=0.4$; $1.1\lesssim K\lesssim1.4$
for $E_{0}=0.5$ etc. In these regimes only one solitary pattern may be formed by the master mode due to MI, and beyond
which many modes will be excited and saturated by the wave number $K$, as can be seen from following simulation results.%
\begin{figure}[tbp]
\begin{center}
\includegraphics[width=4.0in,height=2.5in,trim=0.0in 2in 0in 3in]{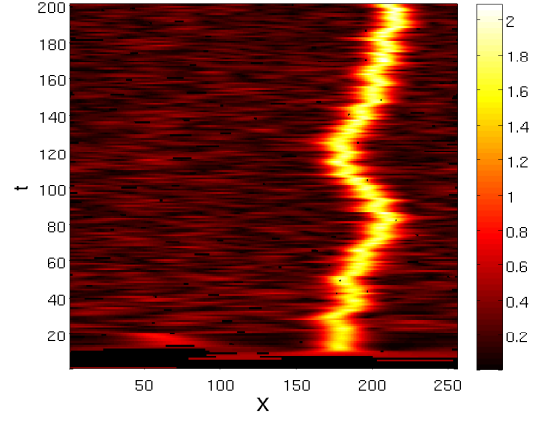} 
\end{center}
\caption{(Color online) Contour of $|E(x,t)|=$const for $K=1.1$ showing that
the pattern selection leads to only one harmonic pattern. The syetm
is in the coexistence of SPC and TC.}
\end{figure}
\begin{figure}[tbp]
\begin{center}
\includegraphics[width=4in,height=3in,trim=0.0in 3in 0in 3in]{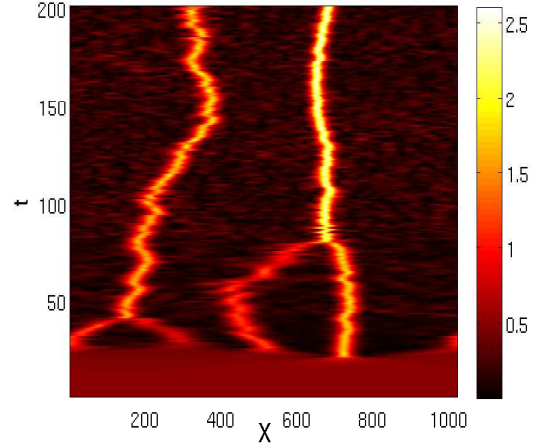} 
\end{center}
\caption{(Color online) Contour of $|E(x,t)|=$const for $K=0.52$. Four solitary
patterns collide pairwise at $t\thickapprox45$
and $t\thickapprox90$ to fuse into two new incoherent patterns. The
syetm is still in the coexistence of TC and STC.}
\end{figure}

We numerically solve the Eqs. (\ref{eq:b1}) and (\ref{eq:b2})
by a fourth-order Runge-Kutta scheme where the spatial derivatives
are approximated with centered second-order diference approximations.
In the numerical scheme we consider the time step $dt=10^{-4}$ and
assume spatial periodicity with the simulation box length $L$$_{x}=2\pi/K$
(the resonant wavelength) together with the grid size $1024$ or $2048$
depending on $K$, so that $x=0$ corresponds to the grid position
$512$ or $1024$. We choose the initial conditions as \cite{STC-He,STC-QZEs,STC-ZEs}
$E\left(x,0\right)=E_{0}+E_{1}\cos\left(Kx\right),\: N=N_{1}\cos\left(Kx\right)$,
 where $E_{1}=b(b_{1}+b_{2}K^{2}),$ $N_{1}=-2bb_{3}E_{0}K^{2}$ and
$b=1/500$ is an arbitrary constant to ensure that the perturbation
is small. Basically, for $K_{c}/2<K<K_{c},$ one unstable mode with
spatially modulational scale $L$$_{x}=2\pi/K$ is excited and then
saturated to form few spatially periodic solitary patterns. However,
as $K$ is lowered from $K_{c}/2,$ there may exist many solitary
patterns with spatial length scales $l_{m}=L_{x}/m,$ where $m=1$
is for the master mode and $m=2,\,3,...,\, M$ for the unstable harmonic
modes with $M<M_{p}=\left[K_{c}/K\right]$ being due to the pattern
selection. Thus, the envelope $E$ can be expressed as \cite{STC-He,STC-QZEs,STC-ZEs,TC-PRL}
\begin{figure}[tbp]
 \begin{center}
\includegraphics[width=4in,height=3in,trim=0.0in 3in 0in 3in]{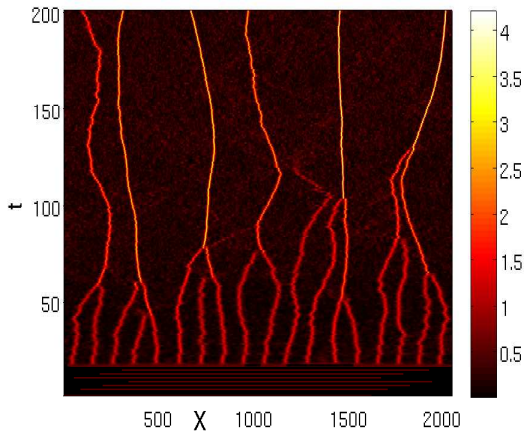} 
\end{center}
\caption{(Color online) Contour of $|E(x,t)|=$const for $K=0.1$. Eighteen
solitary patterns are initially formed from the pattern slection.
After sometime, they collide and fuse to form only six new incoherent
patterns. The Alfv\'{e}n wave emission also occurs everywhere. The
collision is random and not confined between two patterns. The STC
state  emerges. }
\end{figure}
$E=\sum_{m=1}^{M}E_{m}(t)\exp\left(imKx\right)+\sum_{m=M+1}^{\infty}E_{m}(t)\exp\left(imKx\right)$,
 in which the first summation comes from the master mode and the unstable
harmonic modes, whereas the second one is due to the nonlinear interaction
of hf and lf modes. We now proceed with the numerical simulation of
Eqs. (\ref{eq:b1}) and (\ref{eq:b2}) considering $E_{0}=0.5$ for
which the domain of MI for $K$ is $1.1\lesssim K\lesssim1.4$. In the region
$K_c/2=0.7<K\lesssim1.4$, the motion of the solitary pattern is either temporal
recurrent (periodic) or pseudorecurrent (quasiperiodic) depending
on the values of $K$. 

Figure 1 shows that for $K=1.1$, only one master
pattern is formed at $x\thickapprox180$ implying that the system
is stable. However, as can be seen from Fig. 2 that if $K$ is lowered,
i.e., for $K=0.52<0.7$, one master pattern and one harmonic pattern
whose initial peaks are at $x\thickapprox750$ and $550$ exhibit
stochastic behaviors, which results in the coexistence of temporal
chaos (TC) and spatially partial coherence (SPC). It has been
shown that the resonant overlapping may be the cause for TC \cite{TC-PRL}.
Furthermore, in Fig. 2, the pattern selection leads to the excitation of four solitary
modes formed initially from the master mode and unstable harmonic
modes. At $t\thickapprox45$, two solitary patterns initially peaked
at $x\thickapprox50$ and $350$ collide and fuse to form a new pattern
with stronger amplitude. At the same time strong Alfv\'{e}n wave
(AW) emission takes place everywhere and solitary structures are distorted.
As a result, at $t\thickapprox90$, one solitary pattern peaked at
$x\thickapprox550$ collides with the master pattern and gets fused into
another new one. Thus, after the two collisions there remains only
one new incoherent pattern together with the distorted master pattern.
However, as will be evident from the analysis of power spectra (Figs. 4 to 6 below) that 
few collisions are not sufficient for the cause of STC. The coherence
of the system is still partially retained, so that the system is in
the state of TC and SPC. %
\begin{figure}[tbp]
\begin{center}
\includegraphics[width=4in,height=4.0in,trim=0.0in 4in 0in 2in]{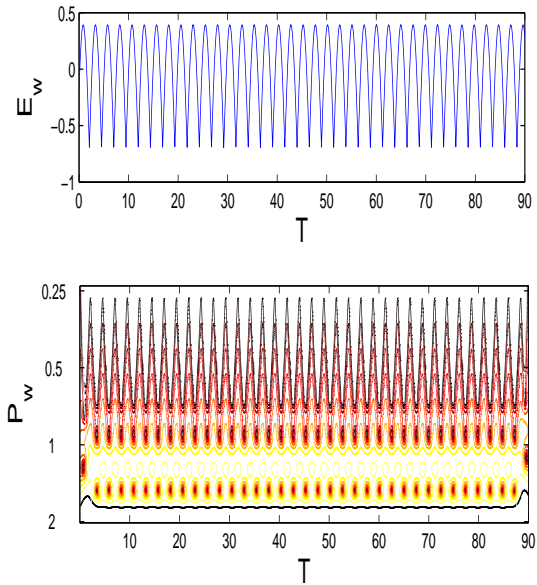} 
\end{center}%
\caption{(Color online) The normalized wave electric field ($E_w$, upper panel)rescaled with the standard deviation of the sampling data and the wavelet power spectra ($P_w$ in logarithmic scale, lower panel) with respect to the sampling time $T$ for $K=1.1$ indicating periodic wave trains.}
\end{figure}

Next, we consider the case in which many unstable modes are excited and 
saturated from the master mode, unstable harmonic modes as well as due to
nonlinear interactions to form many solitary patterns. As for example,
for $K=0.1$, the pattern slection leads to eighteen solitary modes (Fig.
3) in which the fisrt collision occurs at $t$ $\approx45$. At longer
times only few new incoherent patterns remain after collision and
fusion among them due to strong AW emission. Four patterns initially
peaked at $x\thickapprox1700,\,1800,\,1900,\,2000$ collide pairwise
at $t\thickapprox65$ and $80$ respectively to form two new patterns,
which again collide to each other due to strong AW emission at $t\thickapprox130$
to fuse into another new pattern. Also, the harmonic pattern which was initially excited at $x\thickapprox1200$
disappears after some time $t\thickapprox110$ due to AW emission. As time goes on, several
other collisions and fusions take place repeatedly. The patterns are
then much distorted and the original eighteen solitary waves are finally
fused into six new incoherent patterns. 

We now perform a wavelet analysis of the electric field time series
using a Morlet transform. Wavelets have  been widely used for the analysis
of time-frequency spectra in diverse applications including dynamical
analysis of structural systems, image processing as well as pattern
recognition in spatiotemporal systems \cite{Wavelet}. In order to
recognize the spatiotemporal features of the present system as observed in
Figs. 1 to 3, we analyze the corresponding data of the electric field
for different values of $K=1.1,$ $0.52$ and $0.1$ as above. To this end,
we rescale the absolute value of the electric field amplitude as $(E_{W}$)
by the standard deviation of the sampling data and plot against the
sampling time $T$ (with a time step $0.01$) which depends on the
size of the data considered. In our analysis we have used a Morlet
wavelet as the mother wavelet which offers a good balance between
time and frequency localizations and consists of a plane wave modulated
by a Gaussian wave function. The Morlet transform is then used to calculate
the wavelet for the field at the significance level with variance, $\sigma^{2}=1.$
The corresponding wavelet power spectra of the time series shown in
the upper panels of Figs. 4 to
6 are visualized in the lower panels of the same. We see that in a certain interval of $T$, the pattern shows a
regular simple periodic train of waves for $K=1.1$ (Fig. 4). This
result is consistent with the contour plot as shown in Fig. 1. 
When $K=0.52$, Fig. 5 indicates a quasiperiodic wave trains, i.e., chaotic
in time but may be partially periodic in space as analogous to the behaviors
 depicted in Fig. 2. Finally, strong chaotic motion can be seen
from Fig. 6 for $K=0.1$ which agrees with the features observed in
Fig. 3. 
\begin{figure}[tbp]
\begin{center}
\includegraphics[width=4in,height=4.0in,trim=0.0in 5in 0in 2in]{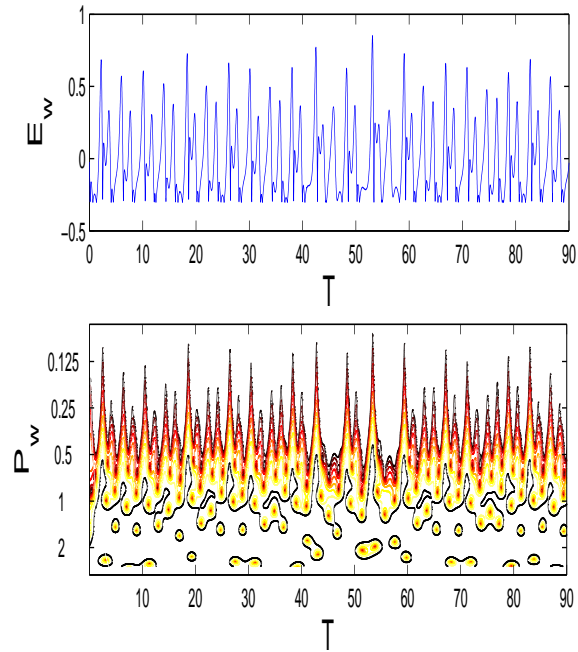} 
\end{center}
\caption{(Color online) The same as in Fig. 4, but for $K=0.52$ indicating the state of temporal chaos but spatially partial coherence.}
\end{figure}
\begin{figure}[tbp]
\begin{center}
\includegraphics[width=4in,height=4.0in,trim=0.0in 4.5in 0in 2.5in]{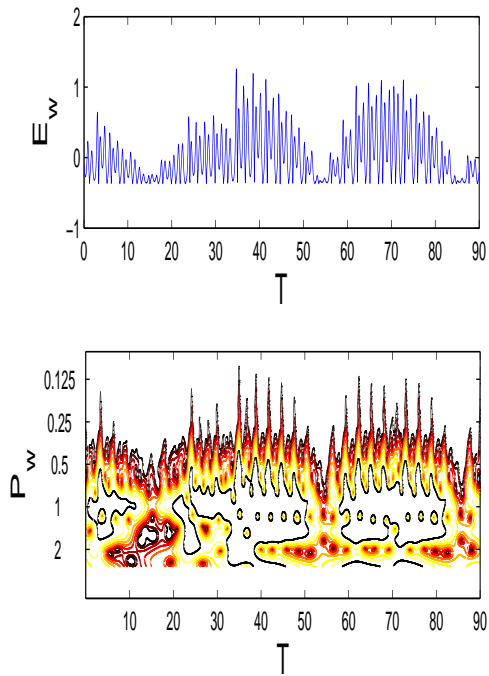} 
\end{center}
\caption{(Color online) The same as in Fig. 4, but for $K=0.1$ indicating that the motion is strongly chaotic in both space and time. }
\end{figure}

Thus, in the process of collision and fusion strong chaotic activity
in both space and time can be observed in the present system. As a result, there remains,
after a longer time, few incoherent patterns together with few stable
higher harmonic modes excited through nonlinear interactions. A certain
amount of the system energy, which was initially confined to many solitary
waves, is now redistributed into those new incoherent patterns as
well as to stable higher-harmonic modes with short-wavelengths $(K>K_{c})$.
The transport of the trapped electrons then occurs, and electrons
can thus be accelerated across the magnetic field via a surfatron
acceleration mechanism \cite{Surfatron}. So, if initially there exist
a number of unstable modulation lengths to form many solitary patterns,
pattern evolution (i.e., collision, fusion and distortion among them)
and strong AW emission can lead to the STC state and hence to the
onset of  turbulence in dusty magnetoplasmas. However, the mechanism that leads to the state of STC 
is still remains unclear. 

We ought to mention that in the present investigation, we have considered the parameter regimes that give positive group dispersion $(\alpha>0)$ of the UH waves. In the opposite case, i.e. for $\alpha<0$, one needs to consider some higher magnetic field strength than that considered here. This may, however, give rise stable wave propagation \cite{UH-Kaufman,Head-on-collision}. Furthermore, the effect of the cubic nonlinear term in  Eq. (\ref{eq:b1}), which appears due to the relativistic electron mass increase, is to enhance the excitation of many more unstable harmonic modes and hence the faster dynamical transition from order to STC, than the nonrelativistic case \cite{Modified-Alfven-Wave}. On the other hand, the stationary charged dust grain, which modifies the frequency of the modulated ES-UHWs and hence the associated instability growth rate, does not change the qualitative behaviors (except the fact that the collision and fusion among the patterns somewhat differ at different times and that the wave amplitude decreases slightly at a specific time) of the nonlinear interactions as shown in Fig. 3. That is, the number of modes initially excited at different modulational length scales and the number of  remaining modes (i.e., six) after collision and fusion among them,  remain the same.  Moreover, the nonlinear coupling of colliding multiple UH wave envelopes (head-on collisions) that may give rise  to  stability of one-dimensional envelope solitons due to negative group dispersion ($\alpha<0$), could be an another important investigation \cite{Head-on-collision}, but is beyond the scope of the present study. 

It is to be mentioned that although, there is no direct or indirect observations, in particular, for localized UH wave envelopes coupled with Alfv{\'e}n waves. However, based on the recent observations, i.e. existence of many electric field solitary structures in the vicinity of Saturn magnetosphere with the magnetic fields $0.1$ nT to $8000$ nT \cite{Solitary-Wave-Saturn}, and also to the fact that the equatorial plane of Saturn contains charged dusts, and dusty plasma can move across the magnetic field lines \cite{Modified-Alfven-Wave}, we conclude that the present results could give an insight for understanding the origin of lf Alfv\'{e}nic turbulence that can occur in  Saturn's magnetosphere. However, as
of 2010, the magnetosphere of Saturn  still remains an important subject of the ongoing investigation by Cassini mission.

APM acknowledges support from the Kempe Foundations,
Sweden, through Grant No. SMK-2647.

\end{document}